\documentclass{ICHD2016}

\newcommand{\be}{\begin{equation}}
\newcommand{\ee}{\end{equation}}

\newcommand{\tx}{\tilde{x}}
\newcommand{\ty}{\tilde{y}}
\newcommand{\tz}{\tilde{z}}

\newcommand{\bOm}{{\boldsymbol \Omega}}

\newcommand{\bq}{\mathbf{q}}
\newcommand{\bde}{\mathbf{e}}
\newcommand{\bk}{\mathbf{k}}
\newcommand{\bv}{\mathbf{v}}
\newcommand{\btr}{\tilde{\mathbf{r}}}

\newcommand{\rme}{\mathrm{e}}
\newcommand{\rmd}{\mathrm{d}}
\newcommand{\rmi}{\mathrm{i}}

\newcommand{\hzeta}{\hat{\zeta}}
\newcommand{\hbv}{\hat{\mathbf{v}}}
\newcommand{\hbOm}{\hat{\bOm}}
\newcommand{\hu}{\hat{u}}
\newcommand{\hv}{\hat{v}}
\newcommand{\hw}{\hat{w}}
\newcommand{\hp}{\hat{p}}

\title{Oscillating sources in a shear flow with a free surface}
\author{\underline{S. \AA. Ellingsen}$^1$ and P. A. Tyvand$^2$}
\address{%
$^1$Department of Energy and Process Engineering,
Norwegian University of Science and Technology,
N-7491 Trondheim, Norway\\
$^1$Department of Mathematical Sciences and Technology,
Norwegian University of Life Sciences,
N-1432 {\AA}s, Norway\\
simen.a.ellingsen@ntnu.no, peder.tyvand@nmbu.no\\
}
\begin{document}
% ---------------------------------------------------------------------
\maketitle
% ---------------------------------------------------------------------

\begin{abstract}
  We report on progress on the free surface flow in the presence of submerged oscillating line sources (2D) or point sources (3D) when a simple shear flow is present varying linearly with depth. Such sources are in routine use as Green functions in the realm of potential theory for calculating wave-body interactions, but no such theory exists in for rotational flow. We solve the linearized problem in 2D and 3D from first principles, based on the Euler equations, when the sources are at rest relative to the undisturbed surface. Both in 2D and 3D a new type of solution appears compared to irrotational case, a critical layer-like flow whose surface manifestation (``wave'') drifts downstream from the source at the velocity of the flow at the source depth. We analyse the additional vorticity in light of the vorticity equation and provide a simple physical argument why a critical layer is a necessary consequence of Kelvin's circulation theorem. In 3D a related critical layer phenomenon occurs at every depth, whereby a street of counter-rotating vortices in the horizontal plane drift downstream at the local flow velocity. 
\end{abstract}

%%%%%%%%%%%%%%%%%%%%%%%%%%%%%%%%%%%%%%%%%%%%%%%
%%%%%%%%%%%%%% S E C T I O N %%%%%%%%%%%%%%%%%%
%%%%%%%%%%%%%%%%%%%%%%%%%%%%%%%%%%%%%%%%%%%%%%%
\section{Introduction}

The use of oscillating sources as Green functions for calculating interactions between surface waves and submerged or floating bodies has been tremendously successful in marine hydrodynamics, describing situations where the flow may be approximated as irrotational and analysed using potential theory \cite{newman77,faltinsen90}. Basic solutions are reviewed, e.g., in \cite{wehausen60}. In the presence of a current which is not spatially uniform, however, these methods are not applicable as they exist today.

We take the first steps towards a corresponding theory in the presence of a shear current. The simplest 2D and 3D cases are considered wherin the sources are presumed to be at rest with respect to the surface (to avoid additional Doppler-related effects), and the shear current is presumed to vary linearly with depth. We linearize all equations with respect to the perturbations caused by the source, which allows superposition of the final solutions to arbitrary distributions of sources. 

Tyvand \& Lepper\o d \cite{tyvand14} considered the linearized 2D problem of an oscillating line source under the assumption that potential theory could be used for the perturbation, a notion motivated by the Kelvin circulation theorem. Their procedure turns out to be flawed, however, because the prerequisites for vorticity conservation are broken at the singular source point, and vorticity is not conserved for particles passing through this point. We here take a step back and consider both the 2D and the more general 3D point source problem from first principles, based on the Euler equations and inhomogeneous continuity equation. An additional far-field solution is then found downstream of the source even in 2D, corresponding to an undulating vorticity sheet directly downstream of the source. The presence of this vorticity perturbation, advected downstream with the flow, may be understood based on the vorticity equation, or with a simple physical argument using the Kelvin circulation theorem.

We present further details of the velocity field and surface waves from a line source and point source, with particular emphasis on the different contributions to the far-field solutions. Comments on additional challenges in using our solutions as Green functions for modelling bodies are given in the Conclusions.

%%%%%%%%%%%%%%%%%%%%%%%%%%%%%%%%%%%%%%%%%%%%%%%
%%%%%%%%%%%%%% S E C T I O N %%%%%%%%%%%%%%%%%%
%%%%%%%%%%%%%%%%%%%%%%%%%%%%%%%%%%%%%%%%%%%%%%%
\section{Mathematical formulation and solution}

\begin{figure}[tb]
  \begin{center}
    \includegraphics[width=.6\textwidth]{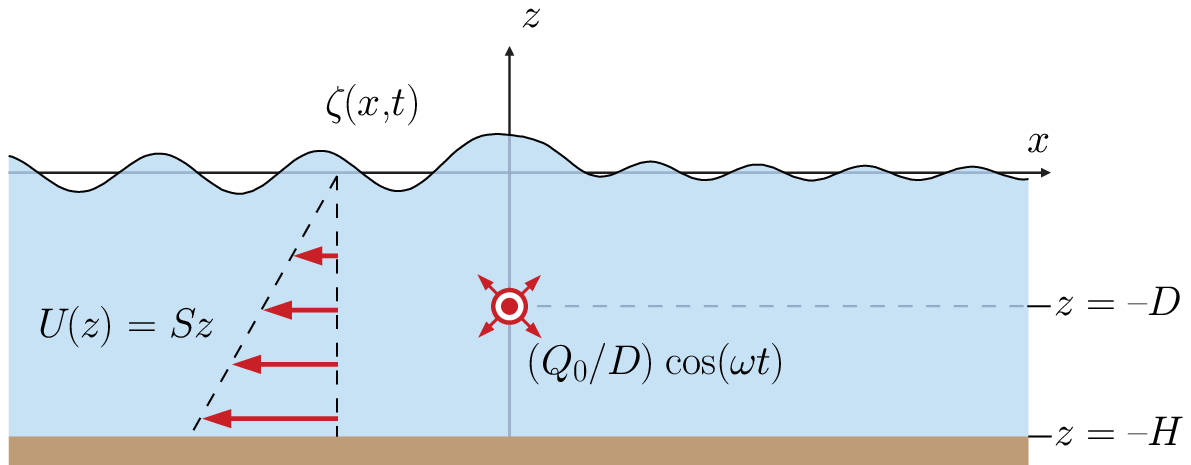}~
    \includegraphics[width=.35\textwidth]{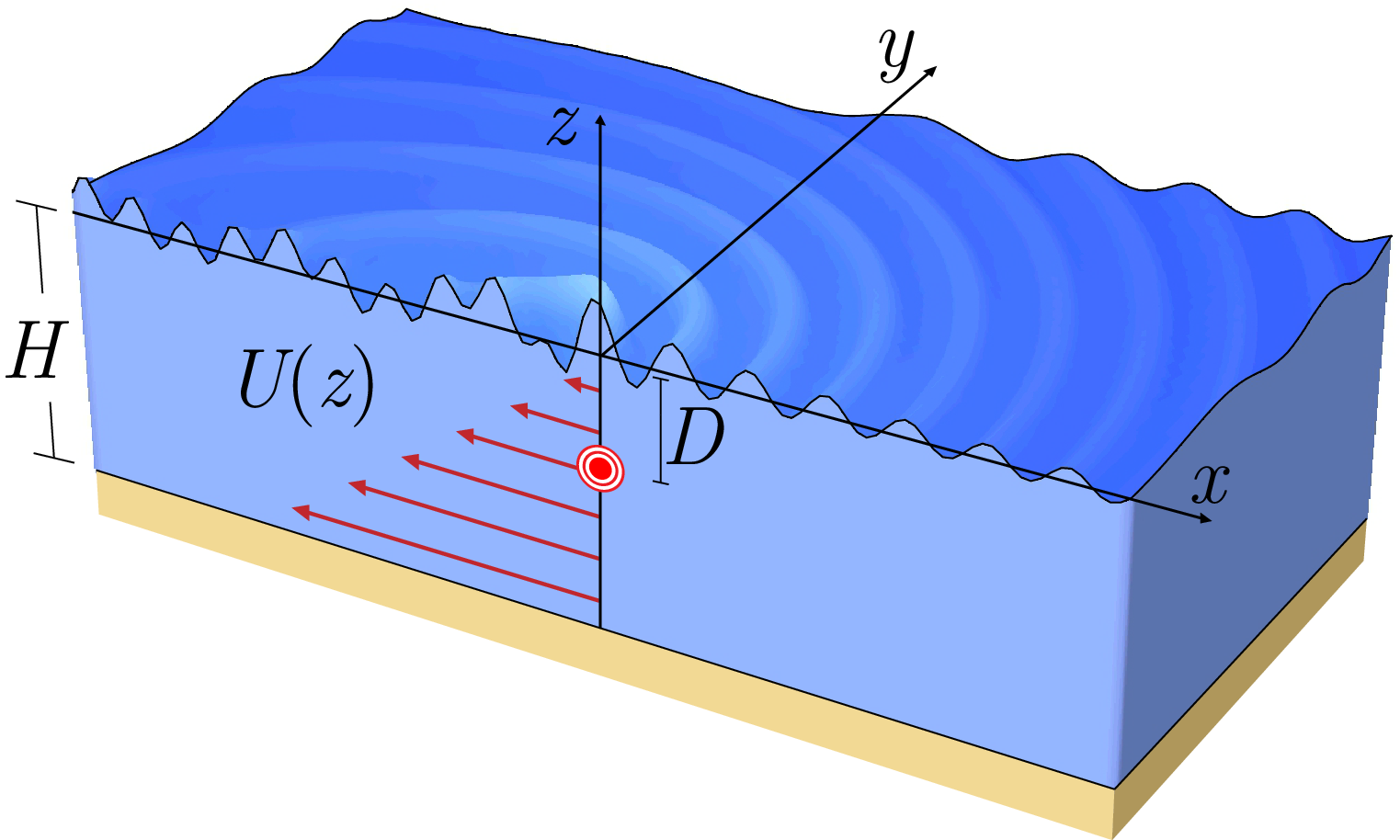}
    \caption{Geometries considered: a source oscillating at frequency $\omega$ a distance $D$ below a free surface, so that the source is at rest relative to the undisturbed surface. Left: A line source (2D), Right: point source (3D). }
    \label{fig1}
  \end{center}
\end{figure}

We consider an oscillating source sitting a depth $D$ below the free surface of a background shear flow whose depth dependence is $U(z) = Sz$, so that the background vorticity is $S\bde_y$ ($\bde_{i}$ is unit vector in $i$ direction with $i\in\{x,y,z\}$). In the 2D set-up, the source is a line source along the line $(x,z)=(0,-D)$, and in the 3D set-up it is a point source at position $(0,0,-D)$. Let the perturbed velocity field be $U(z)\bde_x+\hbv=(U+\hu,\hv,\hw)$ and the position of the free surface relative to the flat state be $\hzeta(x,y,t)$. The pressure is $-\rho g z + \hp$ (zero atmospheric pressure assumed), and the vorticity is $S\bde_y+\hbOm$. Equations of motion and boundary conditions are linearized with respect to perturbation quantities. We assume incompressible flow and neglect the effect of viscosity as well as surface tension. The geometries considered are shown in Fig.~\ref{fig1}.

Now let the tag ``2D'' refer to the line source geometry, and ``3D'' to the point source.
The line and point sources have amplitudes (maximum volume flux) $Q_0/D$ and $Q_0$, respectively, and oscillate with fixed frequency $\omega$, so the continuity equation reads
\be\label{cont}
  \nabla\cdot\hbv = (Q_0/D)\delta(x)\delta(z+D)\rme^{-\rmi\omega t} \text{ (2D) or }
  \nabla\cdot\hbv = Q_0\delta(x)\delta(y)\delta(z+D)\rme^{-\rmi\omega t} \text{ (3D).}
\ee
In the 2D case we let $\partial/\partial y=0$.
We proceed to finding the solution to the linearized Euler equation with ditto boundary conditions,
\be\label{euler}
  \left(\frac{\partial }{\partial t}+U(z)\frac{\partial}{\partial x}\right)\hbv+ S\hw \bde_x = -\frac1\rho\nabla \hp.
\ee
The linearized kinematic and dynamic boundary conditions are
\be
  \hw|_{z=0}=\frac{\partial \hzeta}{\partial t} \text{ and } -\rho g \hzeta +  \hp|_{z=0} = 0.
\ee

 We introduce the Fourier transform in horizontal dimensions:
\be\label{fourier}
  [\hbv,\hp,\hbOm,\hzeta]= \frac{Q_0}{D^2}\begin{cases}\int_{-\infty}^\infty\frac{\rmd q_x}{2\pi} [\bv(\tz,q_x),p(\tz,q_x),\bOm(\tz,q_x),B(q_x)]\rme^{\rmi q_x \tx-\rmi\omega t} & (2D),\\ 
  \int \frac{\rmd^2 q}{(2\pi)^2} [\bv(\tz,\bq),p(\tz,\bq),\bOm(\tz,\bq),B(\bq)]\rme^{\rmi\bq\cdot \btr-\rmi\omega t} & (3D).
  \end{cases}
\ee
where $\btr=(\tx,\ty)$. We introduce nondimensional quantities as given in Table \ref{tab1} by referring to length scale $D$ and time scale $\omega^{-1}$, except for perturbation quantities for which the relevant length scale is $\zeta_0$ (see Table \ref{tab1}). The nondimensional wave vector is $\bq=(q_x,q_y) = (q\cos\theta,q\sin\theta)$. The linearized system is thus described by the nondimensional system parameters $\sigma, h$ and the (squared) Froude number $F^2$.

\begin{table}
  \begin{center}
    \begin{tabular}{ll|ll}
    \hline
    Dimensional &Non-dimensional &Dimensional &Non-dimensional \\
    \hline
    $S$ &$\sigma$=$S/\omega$ &$H$ & $h=H/D$ \\
    $\hzeta$ & $\zeta=\hzeta/\zeta_0$ & $x,y,z$& $(\tx,\ty,\tz)=(x,y,z)/D$\\
    $t$& $T=t\omega$& $\bk$& $\bq=D\bk$ \\
    $g$ & $F = \omega \sqrt{D/g}$ & $Q_0$ &$\zeta_0=Q_0\omega/(g D)$\\
    \hline
    \end{tabular}
    \caption{List of dimensional and corresponding dimensionless quantities. }\label{tab1}
  \end{center}
\end{table}

In Fourier form, the Euler equation (\ref{euler}) and continuity equation (\ref{cont}) can be reduced to the Rayleigh equation for $w$ alone: 
\be\label{rayleigh}
  w''-q^2 w  = \delta'(\tz+1)+\frac{q}{\tz (q-q_{zc})}\delta(\tz+1)
\ee
for 3D, where 
\be
  q_{zc} = (\tz \sigma\cos\theta)^{-1}.
\ee 
The 2D equivalents of Eq.~\eqref{rayleigh} and $q_{zc}$ are found with the replacement rule 
\be\label{replacement}
  q\to q_x ~\text{ and }~ \theta\to 0; ~q_y\to 0.
\ee
We write the Rayleigh equation in the form \eqref{rayleigh} to emphasize that the inhomogeneous term proportional to $\delta(\tz+1)$ introduces a pole singularity at $q=q_{zc}$. A physical understanding of this pole is discussed below. The closeness between 2D and 3D formalisms is even more explicit if we were to express 3D quantities in terms of $q_x,q_y$, but it was found to be preferable to work with polar coordinates in the $\bq$ plane. In the following the 3D case will often be used for reference, with the understanding that the 2D equivalent can be found with Eq.~\eqref{replacement}.

%%%%%%%%%%%%%%%%%%%%%%%%%%%%%%%%%%%%%%%
%%%%%%%%%%% S E C T I O N %%%%%%%%%%%%%
%%%%%%%%%%%%%%%%%%%%%%%%%%%%%%%%%%%%%%%
\section{Formal solution: velocity field and surface elevation}

We proceed to solving the system of equations, first finding $w(\tz,\bq)$ from Eq.~\eqref{rayleigh} and inserting this solution into Eqs.~\eqref{cont} and \eqref{euler} to find remaining velocity components and the surface elevation. In the following the formal solutions are quoted, to be further discussed in subsequent sections.
%\begin{subequations}
%\begin{align}
%  u(\tz,\bq)=& \frac{\rmi q_x}{q}\left\{A(\bq)\cosh q(\tz+h) +\left[\sinh q(\tz+1)-\frac{\cosh q(\tz+1)}{q-q_c}\right]\Theta(\tz+1)+\frac{q_y^2}{q_x^2\tz}\frac{w(\tz,\bq)}{q-q_{cz}}\right\}, \\
%  v(\tz,\bq)=& \frac{\rmi q_y}{q}\left\{A(\bq)\cosh q(\tz+h) +\left[\sinh q(\tz+1)-\frac{\cosh q(\tz+1)}{q-q_c}\right]\Theta(\tz+1)-\frac{1}{\tz}\frac{w(\tz,\bq)}{q-q_{cz}}\right\}, \\
%  w(\tz,\bq)=& A(\bq)\sinh q(\tz+h) +\left[\cosh q(\tz+1)-\frac{\sinh q(\tz+1)}{q-q_c}\right]\Theta(\tz+1).
%\end{align}
%\end{subequations}

\newcommand{\ba}{\left(\begin{array}{c}}
\newcommand{\ea}{\end{array}\right)}
\newcommand{\baa}{\begin{array}{c}}
\newcommand{\eaa}{\end{array}}

%%%%%%%%%%%%%% S E C T I O N %%%%%%%%%%%%%%%%%%
\subsection{Velocity field and surface elevation: formal solutions}

After much tedious algebra, we find it most instructive to write the point source (3D) velocity field in the following form\begin{align}\label{vels}
  &\ba u(\tz,\bq)\\v(\tz,\bq)\\w(\tz,\bq) \ea = \ba\rmi \cos\theta \\ \rmi\sin\theta\\1 \ea \left\{ A(\bq) \baa \cosh\\ \cosh\\ \sinh \eaa q(\tz+h)\right.\notag \\
  &\left.+ \left[\baa \sinh\\ \sinh\\ \cosh\eaa q(\tz+1)-\frac1{q-q_c}\baa \cosh\\ \cosh\\ \sinh \eaa q(\tz+1) \right]\Theta(\tz+1)+\frac1\tz \ba\tan^2\theta \\ -1 \\ 0 \ea\frac{w(\tz,\bq)}{q-q_{cz}} \right\}.
\end{align}
Here,
\be
  A(\bq) = \frac1{\Gamma(\bq)}\left[\left(F^2 + \frac{q-\sigma F^2\cos\theta}{q-q_c}\right)\frac{\sinh q}{\sinh qh}-\left(q-\sigma F^2\cos\theta+\frac{F^2}{q-q_c}\right)\frac{\cosh q}{\sinh qh}\right]
\ee
and
\begin{align}
%   q_{cz}=& q_{cz}(\tz) = 1/(\tz\sigma\cos\theta) \\
   \Gamma(\bq) =& q - F^2(\coth qh+\sigma\cos\theta),\\
   q_c =& -1/(\sigma\cos\theta) = q_{cz}|_{\tz=-1}.
\end{align}

The 2D line source velocity field can again be found using the replacement rule \eqref{replacement}; note that the last term of Eq.~\eqref{vels}, containing the pole at $q=q_{cz}$, vanishes in the 2D case, and that $q_c\to -1/\sigma$ in that case. One may show that the rule \eqref{replacement} gives the same velocity field as simply letting $q_y=0$. This is physically obvious since a linearized line source can be made from a distribution of linearized point sources, amounting to integrating the point source solutions over all $y$. The integral $\int_{-\infty}^\infty \rmd y \,\exp(\rmi \bq \cdot\btr)=2\pi \delta(q_y)$ now dictates this conclusion.

Expression \eqref{vels} is not very useful for numerical purposes when $\tz>-1$, because the terms in the square brackets each diverge rapidly as $q$ (or $q_x$ for 2D) grows large. Using the linearized boundary conditions at the free surface we find that we may write instead
\begin{align}
  \ba u(\tz,\bq)\\v(\tz,\bq)\\w(\tz,\bq) \ea =& \ba\rmi \cos\theta \\ \rmi\sin\theta\\1 \ea \left\{ \rmi\omega B(\bq) \left[\left(\sigma\cos\theta-\frac{q}{F^2}\right)\baa \cosh\\ \cosh\\ \sinh \eaa q\tz-\baa \sinh\\ \sinh\\ \cosh \eaa q\tz\right]\right.\notag \\
  &\left.+\frac1\tz \ba\tan^2\theta \\ -1 \\ 0 \ea\frac{w(\tz,\bq)}{q-q_{cz}} \right\}~\text{ for }\tz>-1
\end{align}
where
\be\label{B}
  \rmi\omega B(\bq) = \frac{F^2}{\Gamma(\bq)\sinh qh} \left[\cosh q(h-1)+\frac{\sinh q(h-1)}{q-q_c}\right]
\ee
and the 2D expression found with Eq.~\eqref{replacement}. 

Incindentally, $B(\bq)$ in Eq.~\eqref{B} is the Fourier transformed surface elevation as defined in Eq.~\eqref{fourier}.

%%%%%%%%%%%%%% S E C T I O N %%%%%%%%%%%%%%%%%%
\subsection{Radiation condition and far-field solution}\label{sec:rad}

\begin{figure}[tb]
  \begin{center}
    \includegraphics[width=\textwidth]{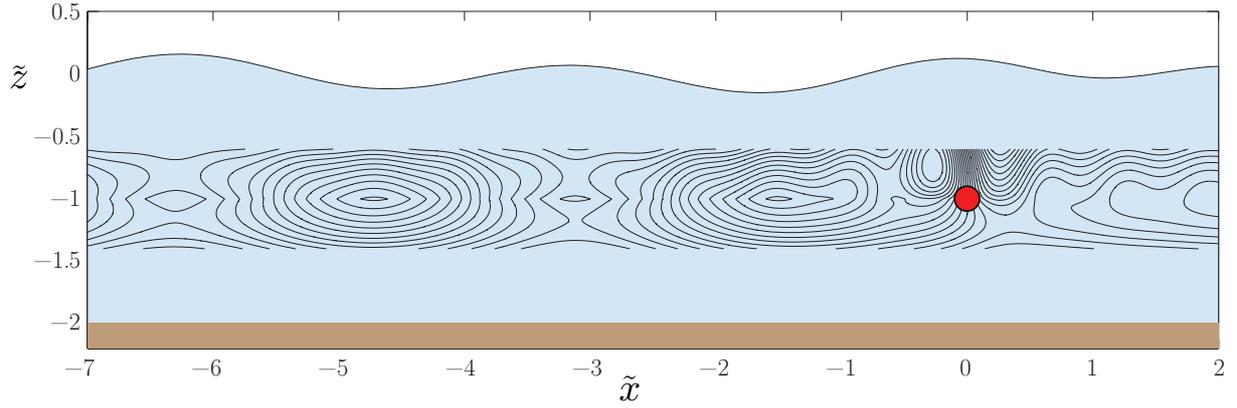} 
  \end{center}
  \caption{Critical layer-like vortex flow downstram of a line source (2D) due to a thin vortex sheet generated by the source combining with the background shear flow. Parameters: $F^2=2,\sigma=0.5,\zeta_0/D=1, h=2,T=0$.}
  \label{fig:critlay}
\end{figure}

The above solutions are ill defined as written because of pole singularities on the axes of integration. This is solved in standard fashion by invoking a radition condition in the manner of, e.g., \S4.9 of \cite{lighthill} by letting $\omega\to\omega(1 + \rmi \epsilon)$ so that $\epsilon\to 0^+$ is taken at the end. 

There are three different kinds of poles present in Eq.~\eqref{vels} and \eqref{B}, and we discuss their physical implications in a moment. For the 3D case we perform the double Fourier integral in polar coordinates, $\int \rmd^2 q = \int_{-\pi}^\pi \rmd\theta\int_0^\infty q\, \rmd q$, so that the $q$ integral is evaluated first. It is well known for such systems (see, e.g., \cite{lighthill} \S4.9 or \cite{wehausen60}) that the waves far from the wave source are given by the contribution to the integrals from these poles only. The pole contribution is calculated by closing the $q$ integral with a contour in the upper (for $\bq\cdot\btr>0$) or lower (for $\bq\cdot\btr<0$) complex plane.

Resonances can occur if two different types of poles coincide, typically resulting in wave elevation and velocity components growing linearly away from the source in the downstream direction. Such resonances, discussed in \cite{ellingsen16b,ellingsen16c}, are beyond the scope of the present article.

%%%%%%%%%%%%%% S E C T I O N %%%%%%%%%%%%%%%%%%
\subsubsection{Regular propagating waves}

The first kind of pole appears where the dispersion relation $\Gamma(\bq)=0$ is satisfied, and is the standard pole to appear for periodically forced wave systems. Only waves satisfying this condition may propagate outwards towards infinity. The terms proportional to $A(\bq)$ in Eq.~\eqref{vels}, as well as the first term in the brackets of Eq.~\eqref{B}, correspond to the homogeneous solutions to the wave problem for velocity field and surface elevation, respectively.

%%%%%%%%%%%%%% S E C T I O N %%%%%%%%%%%%%%%%%%
\subsubsection{Critical layer-like vortex flow}

The second kind of pole is where $q=q_c$ (3D) or $q_x=q_c$ (2D). The form of $q_c$ is different in the two cases, but in both cases the  pole is found where $q_x = -1/\sigma$. Interpretation is easier in dimensional units, where we may write
\be
  \lambda_x\equiv \frac{2\pi}{k_x} = \frac{2\pi}{\omega}U(-D) ~~\text{ or }~~ \frac{\omega}{k_x}=U(-D).
\ee
In other words, this pole corresponds to a wave whose wavelength in the $x$ direction equals the distance travelled by a fluid particle advected with velocity $U(-D)$ (i.e., the velocity at the source's depth) during one oscillation period $2\pi/\omega$. Indeed we show below that a train of vortical flow structures similar to a critical layer will exist directly downstream of the oscillating source. The source injects a periodical vorticity perturbation which is transported downstream with the flow, whose spatial period is $2\pi U(-D)/\omega$. This becomes clearer when the vorticity equation is discussed below, and we also argue how this is a necessary consequence of Kelvin's circulation theorem. The $q=q_c$ pole in $B$ corresponds to a non-dispersive ``wave'' travelling downstream at velocity $U(-D)$, and is the surface manifestation of the vortical flow. The phase velocity $\omega/k_x$ equals the velocity at the source depth, for which reason we refer to it as a ``source critical layer''. For the point source this far-field wave is restricted to a thin strip downstream of the source, and for both 2D and 3D the pole exists on the downstream side only. 

We discuss the source critical layer further below in the context of the vorticity equation. Much further discussion of the source critical layer phenomenon will be found in forthcoming publications \cite{ellingsen16b,ellingsen16c}. 

%%%%%%%%%%%%%% S E C T I O N %%%%%%%%%%%%%%%%%%
\subsubsection{Critical layer in horizontal velocity}

The horizontal velocity components $u,v$ also have a third kind of pole, where $q=q_{cz}$. Since this pole does not exist in the $w$ component, its corresponding far-field contribution is not seen in the surface elevation $B$. Moreover, it also only appears for $3D$; there is no 2D equivalent. Similarly to the previous case, $q=q_{cz}$ may be written
\be\label{crithor}
  \lambda_x\equiv \frac{2\pi}{k_x} = \frac{2\pi}{\omega}U(z) ~~\text{ or }~~ c_x\equiv\frac{\omega}{k_x}=U(z).
\ee
For any depth $z$ in the fluid domain there exists a value of $k_x$ so that the phase direction in the $x$-direction equals the flow velocity at that depth. This is the criterion for a critical layer. Since wave vectors from the point source have all directions, and $k_x$ takes all values between $-k$ and $k$ as $\theta$ is varied, Eq.~\eqref{crithor} can always be satisfied for some $q$ in downstream directions ($q_x<0$), no matter what $z$ is. Also this pole can only occur for downstream-pointing $\bk$, since $q_{cz}<0$. 

\begin{figure}[tb]
  \begin{center}
    \includegraphics[width=\textwidth]{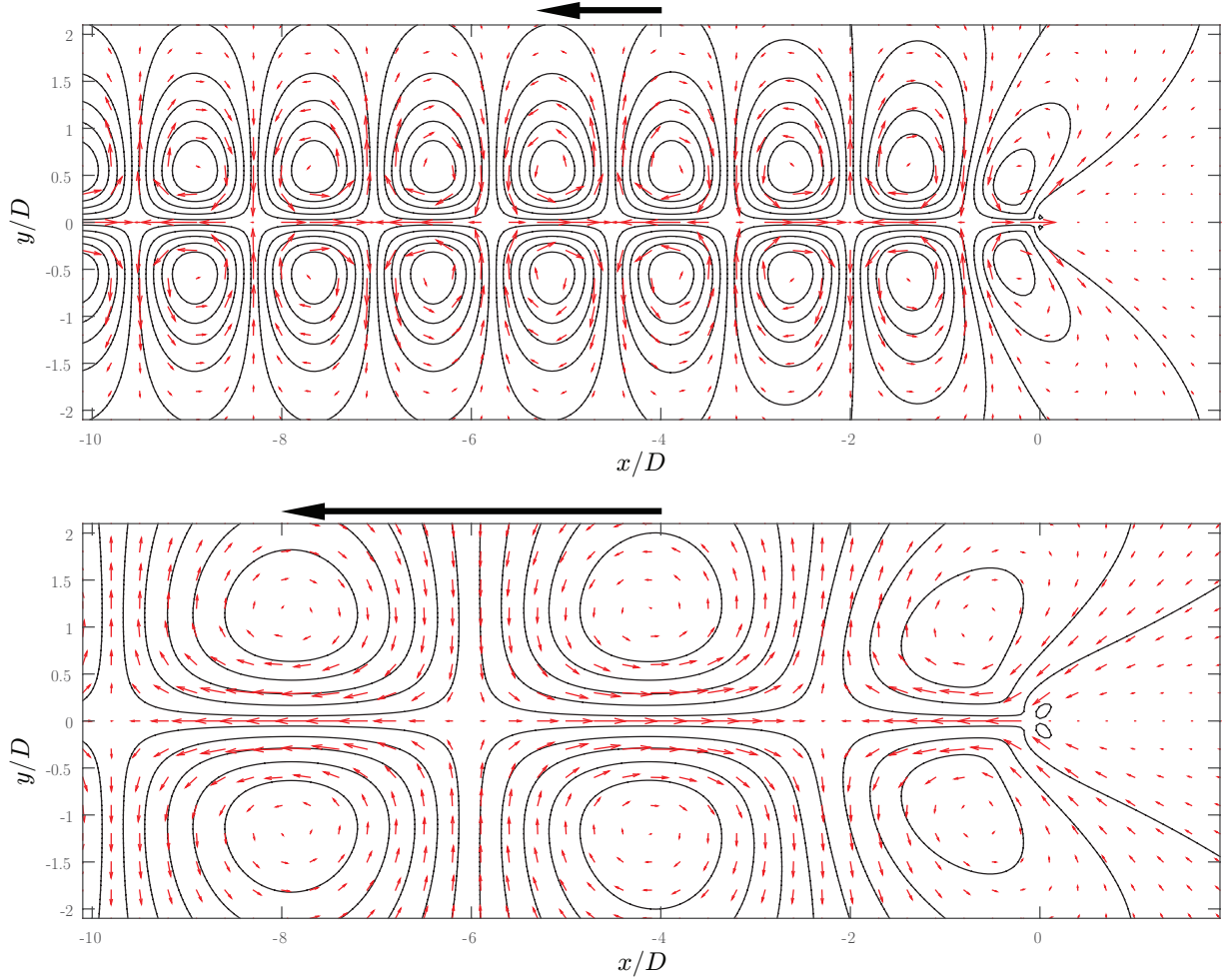} 
  \end{center}
  \caption{Far-field velocity components $\hu,\hv$ from the point source in the horizontal plane due to the pole at $q=q_{cz}$, corresponding to a critical layer-contribution at every level $\tz$. $\tz=0.5$ (top) and $1.5$ (bottom). Parameters are $F^2=1, \sigma=0.8, h=\infty, T=0.3$. The vortex-like structures each drift downstream at velocity $U(z) = Sz$ as indicated by the black arrows.}
  \label{fig:vels}
\end{figure}

In Figure \ref{fig:vels} we plot, for the first time, the far-field contribution to horizontal velocity $(\hu,\hv)$ due to the pole at $q=q_{cz}$, at two different values of $\tz$. The flow pattern that emerges is a symmetric street of counter-rotating vortices which drift downstream with the background flow's velocity $U(z)$. Fig.~\ref{fig:vels} shows one plane sitting between source and free surface ($\tz=0.5$) and another positioned below the source ($\tz=1.5$). Note that the vortical flow structures nearest the source (at $\tx=\ty=0$) rotate in opposite senses. The black lines are streamlines in the horizontal plane, arrows indicate velocity direction and magnitude. The wavelength of periodicity is $\lambda_x(\tz)$, proportional to $|\tz|$ for the Couette profile.

%%%%%%%%%%%%%%%%%%%%%%%%%%%%%%%%%%%%%%%%%%%%%%%
%%%%%%%%%%%%%% S E C T I O N %%%%%%%%%%%%%%%%%%
%%%%%%%%%%%%%%%%%%%%%%%%%%%%%%%%%%%%%%%%%%%%%%%
\section{Vorticity perturbations}

As seen above, both the line source and the point source combine with the shear flow to create rotating downstream flow structures \emph{in addition} to regular waves propagating away. In the case of the line source in particular, this may be a surprising fact in light of Kelvin's circulation theorem which implies that the vorticity of each fluid particle is preserved in strictly 2D flow. Indeed we argue below that the additional vorticity can be seen as a consequence of Kelvin's theorem, and stems from the fact that continuity is broken in the single point where the source is positioned. 

It is most instructive to consider the linearized vorticity equation for the system at hand. For maximum clarity we will discuss the 2D case first, then the slightly more involved 3D case, even though the former is (as usual by now) included in the latter.

\newcommand{\rmD}{\mathrm{D}}
In the 2D problem the vectorial vorticity equation in dimensional units, linearized with respect to perturbation quantities, reads
\be\label{vort2D}
  \frac{\rmD\hbOm}{\rmD t}= - S (\nabla\cdot\hbv) \bde_y = - \frac{SQ_0}{D}\delta(x)\delta(z+D)\rme^{-\rmi \omega t}
\ee
where we have inserted continuity equation \eqref{cont} in the last form. Eq..~\eqref{vort2D} shows that vorticity is conserved for a fluid particle \emph{except} if it travels through the point $(x,z)=(0,-D)$, in which case it will pick up an additional vorticity due to the inhomogeneous term in \eqref{vort2D}. The time the particle spends in the vicinity of the source is proportional to $1/U(-z)$, so the additional vorticity found in a thin strip directly downstream of the source could be expected to be proportional to $-SQ_0\exp(-\rmi\omega t)/DU(-z)$. Said additional vorticity will oscillate in sign with a spatial wavelength of $2\pi|U(-z)|/\omega$, the distance traveled by a particle at depth $-D$ during one oscillation period. This we show to be the case below. An example of this additional vorticity is seen in Fig~\ref{fig:critlay} where a discontinuity in the horizontal velocity (i.e., a vorticity sheet) at $z=-D$ is visible downstream of the source.

In the 3D case the linearized vorticity equation for our system becomes
\be\label{vort3D}
  \frac{\rmD\hbOm}{\rmD t}=S\frac{\partial\hbv}{\partial y} - S (\nabla\cdot\hbv) \bde_y =S\frac{\partial\hbv}{\partial y} - SQ_0 \delta(x)\delta(y)\delta(z+D)\rme^{-\rmi \omega t}.
\ee
We find once again the $\delta$-function term which gives an additional vorticity to any fluid particle passing through the source point, resulting in a thin straight line of undulating vorticity directly downstream. We will see below that this thin strip of vorticity has a surface manifestation which we call the ``critical wave''. 
Compared to the 2D version, Eq.~\eqref{vort3D} has an extra term on the right hand side. This term comes from the fact that a wave travelling at an oblique angle with respect to the underlying shear current will shift and twist the vortex lines gently as it passes. The effect was investigated in Ref.~\cite{ellingsen16}, and is related to a theorem that only strictly 2D free surface flow can have spatially constant vorticity \cite{constantin11}.

%%%%%%%%%%%%%% S E C T I O N %%%%%%%%%%%%%%%%%%
\subsection{Vorticity field}

With nondimensional quantities as defined above, the Fourier-transformed vorticity equation (\ref{vort3D}) may be written
\be
  -\rmi\tz \cos\theta(q-q_{cz})\bOm = \rmi q_y \bv + \Omega_z \bde_x - \delta(\tz+1)\bde_y
\ee
where $\bOm=(\Omega_x,\Omega_y,\Omega_z)$. With velocity components already found, the vorticity components are immediately found to be
\begin{subequations}
\begin{align}
  \Omega_x =& \frac{q\tan\theta}{\tz (q-q_{cz})}\left[u(\tz,\bq) -\frac{\rmi  w(\tz,\bq)}{\tz \cos\theta(q-q_{cz})}\right] = \frac{q v(\tz,\bq)}{\tz(q-q_{cz})},\\
  \Omega_y =& \frac{q\tan\theta}{\tz (q-q_{cz})}v(\tz,\bq)-\frac{\rmi }{\cos\theta (q-q_c)}\delta(\tz+1),\label{omy}\\
  \Omega_z =& \frac{q\tan\theta}{\tz (q-q_{cz})}w(\tz,\bq).
\end{align}
\end{subequations}
The last, simpler, form of $\Omega_x$ may be shown from the $x$ and $y$ components of \eqref{euler}.

Notice in particular the last term in component $\Omega_y$ proportional to $\delta(\tz+1)$. In 2D ($\tan\theta=0$) this is the \emph{only} nonzero term, as may be concluded from the vorticity equation \eqref{vort2D} directly. Solving the Fourier integrals using the radiation condition as described, this term, which represents a thin sheet (2D) or line (3D) of vorticity directly downstream of the source, may be found to be
\be\label{omdelta}
  \ba\Omega^\text{(2D)}_y\\ \Omega^\text{(3D)}_{y,\delta}\ea = -\frac{Q_0}{D^2}\rme^{-\tx/\sigma-\rmi T}\ba 1\\ \delta(\ty)\ea \delta(\tz+1)\Theta(-\tx)
\ee
where $\Theta$ is the Heaviside function, and subscript $\delta$ indicates that only the last term of \eqref{omy} is included. 

The full 3D vorticity field has additional contributions which all vanish as $q_y\to 0$; in other words these 3D vorticity perturbations may be recognised as the effect of wave components travelling at oblique angles with respect to the basic shear flow, as discussed in \cite{ellingsen16}.

The presence of the sheet of additional downstream vorticity in 2D may be understood also from the Kelvin circulation theorem. The source does not produce any vorticity in itself, hence any material loop travelling with the flow must retain a constant circulation even if it envelops the source for a period of time as it passes. However, while inside the loop the flux from the source will increase or decrease the mass contained inside it, hence the area encircled must change and in consequence the mean vorticity inside (which is circulation per area). Carrying out this argument formally in 2D gives exactly the expression \eqref{omdelta}. A physically similar process occurs in 3D, where the situation is a little more tangled due to the presence of other vorticity perturbation terms. 

%%%%%%%%%%%%%%%%%%%%%%%%%%%%%%%%%%%%%%%%%%%%%%%
%%%%%%%%%%%%%% S E C T I O N %%%%%%%%%%%%%%%%%%
%%%%%%%%%%%%%%%%%%%%%%%%%%%%%%%%%%%%%%%%%%%%%%%
\section{Surface waves}

\begin{figure}[tb]
  \begin{center}
    \includegraphics[width=\textwidth]{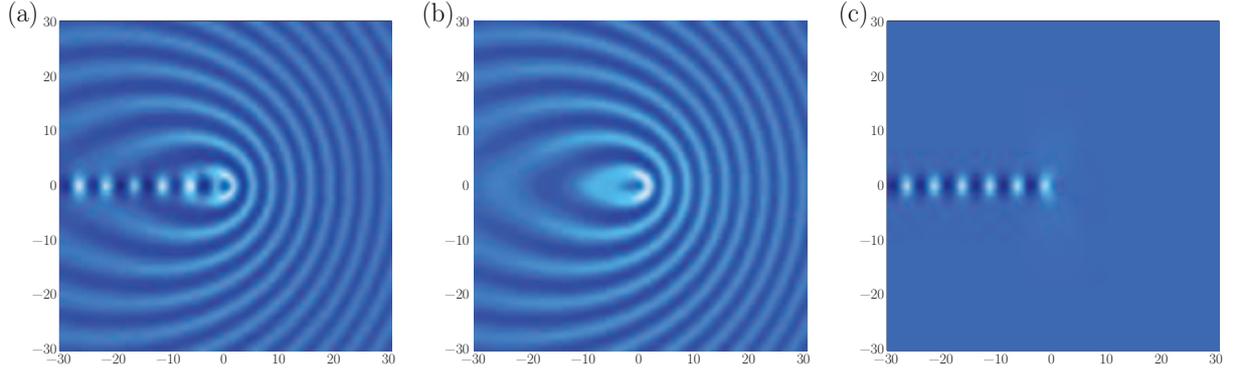} 
  \end{center}
  \caption{Surface perturbations in 3D for sub-resonant parameters $F^2=1, h=4, \sigma=0.8, T=0.3$. The colour scaling of each relief plot is the same, with light colour representing crests and dark troughs. (a) Full surface elevation calculated directly from double Fourier integral with $\epsilon=0.001$. (b) Far-field contribution from regular waves, i.e., the pole where $\Gamma(\bq)=0$. (c) Far-field contribution from the critical layer-like string of downstream vortices, the ``critical wave'', from the pole at $q=q_c$. }
  \label{fig:surf}
\end{figure}

The surface perturbation $\hzeta$ is calculated from the quantity $B$ in Eq.~\eqref{B} and the definition \eqref{fourier}. As discussed above, the integrand $B$ has two different poles, corresponding to two different kinds of far-field waves. The regular, propagating waves which (when shear is moderate) take the form of (skewed) ring waves, correspond to the poles where $\Gamma(\bq)=0$. In addition there is a ``wave'' due to the source critical layer of vortices drifting downstream of the source, as illustrated in the 2D case in Fig.~\ref{fig:critlay}. The horizontal velocity pattern shown in Fig.~\ref{fig:vels}, however, is not visible on the surface. Fig.~\ref{fig:surf} shows an example of the surface waves from a point source. We plot the full (3D) surface elevation in panel (a), calculated with brute force from the Fourier integral using $\epsilon=0.001$ to move the poles slightly off the real axis as described in Sec.~\ref{sec:rad}.  In panel (b) is shown only the contribution from the pole where $\Gamma(\bq)=0$, contributing the far-field regular waves which appear as skewed ring waves. In panel (c) is shown only the critical wave stemming from the sub-surface string of vorticity downstream of the source. The critical wave travels downstream at velocity $U(-D)$, the flow velocity of the source, while the regular waves travel at a different velocity dictated by the dispersion relation $\Gamma(\bq)=0$. %Note that the horizontal vortex motion shown in Fig.~\ref{fig:vels} does not affect the vertical motion, and is thus not visible at the surface.

 The parameters used in Fig.~\ref{fig:surf} are such that the regular-wave and critical wave poles can never co-incide, that is, $\Gamma(q_c,\theta)\neq 0$ for all $\theta$. The sub-resonant case occurs for $F^2$ below a critical resonant value which may be shown to be \cite{ellingsen16c}
\be
  F^2_\text{res} = \min_{\theta, >0}\left\{\sigma\cos\theta[\coth(h\sec\theta/\sigma)-\sigma\cos\theta]\right\}^{-1}
\ee
where the notation means that the minimum value is found with respect to $\theta$, but so that negative values are ignored. When $F^2=F^2_\text{res}$, a resonance occurs between the regular and critical waves, and wave amplitudes directly downstream of the source increase linearly as a function of $|x|$, without bound (in the linear theory) \cite{ellingsen16c}. 

\section{Further discussion and conclusions}

A question which remains open at this point concerns the usefulness of the singular source solution presented above for the purposes of modelling bodies in a shear current. In potential theory this technique has been extremely successful. Essentially, sources and sinks (or continuous distributions thereof) force streamlines in a uniform flow to take detours such as they would around a submerged body, and since in inviscid theory a streamline and a body is formally equivalent, the ability to shape the streamline pattern in this manner provides a way to describe bodies of arbitrary shape interacting with a wave field. Advanced panel methods have been developed for this purpose within the confines of potential theory, cf., e.g., \cite{newman77, faltinsen90}. 

Whether and how source singularities (such as that analysed herein) can be put to the same use in the presence of a shear flow is a question to consider in detail in the future, but some thoughts may be presented already at this stage. Immediately, the thin stream of vorticity advected downstream from the source appears to be a problem: a discontinuous velocity field is thus created, and in 2D the circulation theorem would also disallow a body introducing any additional vorticity into the flow. The problem does not seem unsourmountable, however, because of the ability of a downstream source to absorb the ``source critical layer" vorticity from an upstream one. Two sources aligned in the streamwise direction can thus be brought very close to each other and form a ``detuned'' dipole for which the $\delta$-function term vanishes from the vorticity equations \eqref{vort2D} and \eqref{vort3D}, and no such downstream vorticity results. The detailed properties of the composite singularity thus constructed have yet to be laid out in detail. 

We have studied a basic building block towards a Green function theory for surface waves in the presence of a shear flow, namely a periodically oscillating singular mass source. The problem is solved an analysed both in 2D (line source) and 3D (point source). To keep matters manageable at this stage, the sources are kept at rest relative to the surface so that no Doppler effects appear. A number of striking features are found compared to the still-water potential theory analogue. The least surprising discovery might be that the ring waves emanating from a point source are not circular, an effect already seen previously for the sheared Cauchy-Poisson problem \cite{ellingsen14b}. More striking is the fact that the source, both in 2D and 3D, a thin sheet (2D) or string (3D) of additional vorticity is produced by the source and current and advected downstream. This results in an additional, ``critical'', wave travelling at the current velocity at the source depth. It is termed ``critical'' because its phase velocity $\omega/k$ equals the flow velocity at the source level. In 3D a related critical layer phenomenon is found at \emph{all} depths $z$, manifesting itself in the horizontal velocity components only, where a street of counter-rotating vortices are found downstream, advected at velocity $U(\tz)$ and with spatial period $2\pi |U(z)|/\omega$.

% ---------------------------------------------------------------------

% ---------------------------------------------------------------------

% ---------------------------------------------------------------------
\end{document}